# Introducing Big Data topics: a multicourse experience report from Norway

Beathe Due
Østfold University College
B R A Veien 4,
1783 Halden, Norway
00 47 6921 5000
beathe.due@hiof.no

Monica Kristiansen
Østfold University College
B R A Veien 4,
1783 Halden, Norway
00 47 6921 5000
monica.kristiansen@hiof.no

Ricardo Colomo-Palacios
Østfold University College
B R A Veien 4,
1783 Halden, Norway
00 47 6921 5000
ricardo.colomo-palacios@hiof.no

Dang Ha The Hien
eSmart Systems
Håkon Melbergs vei 16,
1783 Halden, Norway
00 47 451 40 689
dang.hien@esmartsystems.com

## ABSTRACT

In the last few years we have witnessed an explosion of interest in Big Data in both academic and industry arenas. Big Data is about the capture, storage, analysis and visualization of huge volumes of data in both structured and unstructured forms generated from a myriad of applications and devices in a wide set of scenarios. Focusing on the need of academia to develop professional posing the competences that industry demands, the paper presents the approach adopted by the Faculty of Computer Sciences at Østfold University College, Norway to deploy Big Data related contents throughout its studies in the computing field. This paper describes initiatives in bachelor and master programs along with continuous education courses with regards to Big Data topics. New master courses were implemented in the 2014-2015 academic year, while bachelor and continuous education courses will be deployed the year after. Initial results in terms of course assessments and students´ acceptation unveil promising perspectives for the initiative.

## Categories and Subject Descriptors

H.2.8 [**Database Applications**]: Data mining. I.2.2 [**Automatic Programming**] Automatic analysis of algorithms.

## General Terms

Algorithms, Performance, Human Factors.

## Keywords

Big Data; Education.

## 1. INTRODUCTION

According to a report from McKinsey [15], the top priority objective of IT organizations over the next 1-3 years is to "generate more value from data". This objective, lays, among other IT advances on Big Data technologies. According to Forrester, one of the top players in the consultancy arena, Big Data is a key trend for many industries [16], and Gartner, another of the world's leading information technology research and advisory companies, has been pointing out Big Data as one of the strategic technology trends to watch in the next years. Moreover, Gartner states that Big Data is destined to help organizations drive innovation by gaining new and faster insight into their customers [14].

Scientific and commercial literature has provided many definitions for Big Data but agreeing with [20], it means data that is too big, too fast or too hard for existing tools to process. This simplified definition will be extended in the literature review section. In any case, the fact is that due to the data tsunami that comes from the expanding mobile base, internet usage including social networks and sensors, naming just a subset of the data sources; managing and analyzing this information in an accurate way become a critical success factor for organizations around the world.

The Big Data phenomenon is affecting a wide range of fields of study. Thus, computer scientists, physicists, economists, mathematicians, political scientists, bio-informaticists, sociologists, and other scholars are clamoring for access to the massive quantities of information produced by and about people, things, and their interactions [4]. As a result of this, research driven by Big Data reflects a discipline that, to extract meaning from very large datasets, incorporates various techniques such as data mining and visualization into diverse fields [23]. Given the importance of the topic, literature has reported many definitions on the term; one of the most accurate ones is the one provided by [9] in which the term is defined as the set of technologies that comprises datasets that have become too large to handle with the traditional or given computing environment. However, technology world has been facing data challenges from its beginnings; the difference comes in the definition of the term Big, that has been changing from megabytes in the 1970s to the petabytes or Exabytes today [3]. Given that the problem is not new, the origin of Big Data research is rooted in various initiatives started in the early 1970s [23], with the book "Concise Survey of Computer Methods" [22] as the most credited origin of the discipline. However, according to [25], the early 1990s is the beginning of the field of Big Data research. Whatever the case may be, although research based on Big Data can be conducted in various ways, its basic purpose lies in handling huge amounts of data from technological, sociological, and economic systems to discover some hidden patterns [23].

Big Data approaches face two different issues [18]: big throughput and big analytics; the former includes the problems associated with storing and manipulating large amounts of data and the latter those concerned with transforming this data into knowledge. Thus, Big Data solutions require new approaches to obtain insights from highly detailed, contextualized, and rich contents that may require complex math operations, such as machine learning or clustering [5]. These approaches include a variety of applications like business process analytics [27, 28], learning support [13], healthcare [10, 26], government [17] or journalism [19], mentioning just some of the most recent cases.

The remaining of the paper is structured as follows. Section 2 includes insights on Big Data education. Section 3 presents the initiatives launched at the Faculty of Computer Sciences at Østfold University College in the field of Big Data education and, finally, Section 4 wraps-up the paper, introducing main conclusions and suggests futures avenues of research.

## 2. BIG DATA AS A SUBJECT

Agreeing with Agarwal and Dhar [1], big data is possibly the most significant "tech" disruption in business and academic ecosystems since the meteoric rise of the Internet and the digital economy. However, and in spite of its attractiveness, there is a shortage of highly skilled persons for data-related jobs all over the world. Focusing on Europe, this scarcity of talent is threatening the ability to exploit the full potential of Big Data in the zone [12]. The cause may be current times of scarcity of talent in IT [6, 7, 24] or maybe the need of a new set of technical competences. In work environments, demand for Big Data expertise across a range of occupations saw significant growth in 2014 according to Forbes [8]; moreover, in this report, the author reported that there was a 123.60% jump in demand for Information Technology Project Managers with Big Data expertise, and an 89.8% increase for Computer Systems Analysts. Finally, to make the work even more attractive, the median salary for professionals with Big Data expertise is $103,000 a year. Sample jobs in this category include Big Data Solution Architect, Linux Systems and Big Data Engineer, Big Data Platform Engineer, Lead Software Engineer, Big Data (Java, Hadoop, SQL) and others.

According to [21], there is a talent shortfall in data strategy and in a wide variety of technical data management positions largely due to the shortfall in university, professional, and executive education programs designed to produce the talent needed to fill the growing demand for every type of Big Data professional. This could be rooted in the lack of people in academia with relevant knowledge and experience in Big Data environments.

Thus, DBWorld, one of the most important mailing lists for computing, is witnessing an important increase of the importance of Big Data in its posts. Only in the first three months of 2015, the term appeared 73 times in the message subject. Only one of these was repeated, so there were 72 unique postings on Big Data in just 3 months. Almost one per day. Among these 72 postings, 10 were positions announcements including a variety of tenure tracks, lectureships, postdocs and professorships. It is a fact that many universities are developing analytics curricula to educate future data professionals [2], adapting many of their traditional database and business intelligence courses to the new arena of Big Data [5, 21, 29], but also, although in a lesser extent, shifting their Expert Systems or Artificial Intelligence courses to the emergent topic [29]. Finally, Big Data is also beginning to be present in non-computing disciplines like, for instance health [11].

Beyond byzantine discussions on who among computing professionals started off with a comparative advantage with respect to big data [1, 5], the truth is that universities and scholars must integrate fields like computer science, information systems and computer engineering to develop students in the three essential stages of Big Data: Data Collection, Data Analysis and Data Visualization. This paper presents the initiative taken by the Faculty of Computer Sciences at Østfold University College, Norway to integrate Big Data as a topic in its Bachelor, Master and continuous education courses.

## 3. THE COURSES

In this section, authors present the experiences that take place at the Faculty of Computer Sciences, Østfold University College, Norway in three different scenarios: at bachelor level, at master level and in continuous education courses.

### 3.1.1 Bachelor Level

Faculty of Computer Sciences is offering four bachelor programmes, namely: Bachelor's programme in Computer Engineering, Bachelor's programme in Digital Media, Bachelor's programme in Computer Science and, finally Bachelor's programme in Information Systems. Focusing on the last programme, this schedules 180 credits (3 academic years) with 150 credits in compulsory courses and 30 more on electives. From autumn 2015 the course "Large amounts of data: processing and analysis" (10 ECTS) has been scheduled as a substitute to the previous course "Database Administration and Systems". This new course has also lead to the update of the course "Databases" that will include some of the former contents of "Database Administration and Systems", including Brief market and secondary aspects and introductory data warehousing concepts.

Focusing on the course "Large amounts of data: processing and analysis" this course is new for the 2015/16 academic year. Although all concepts could relate to Big Data, strictly speaking not all contents are aiming to cover data fulfilling the "V" characteristics like volume, velocity, variety, value, variability, viscosity and virality. Thus, aspects of traditional advanced bachelor courses in databases like, for instance, database configuration, monitoring & analysis, stored procedures and triggers, replication, query optimization and data warehousing are introduced before covering aspects of data capture, storage, processing and visualization including pure Big Data tools implementation, like, for instance, Hadoop and Hive.

Regarding learning and teaching methods, the course is largely based on a combination of traditional lectures and project work. Some topics included in the projects will not be lectured, and it is up to the students to familiarize themselves with contents in order to be able to apply the knowledge and skills in exams and projects.

Finally, with regards to evaluation, this aspect is highly influenced by teaching approach. Projects (it is scheduled to assign 4 projects per course) represent 49% of the final grade while a written exam is aimed to cover the remaining 51%. This exam is designed to last 3 hours and no resources are allowed. At least one of the 4 projects scheduled are devoted specifically for Big Data topics covered in the course and require implementations on the topic.

### 3.1.2 Master Level

The Faculty of Computer Sciences at Østfold University College is also offering a master degree in "Applied Computer Science" to its students. The master programme in Applied Computer Science extends over a period of two years and awards totally 120 ECTS, of which 45 ECTS credits make up the master's thesis. Courses and course contents were updated in the 2014-2015 academic year. As a consequence of that, a new course entitled "Advanced Topics in Information Systems" (15 ECTS credits) appeared in the program. The course is designed as a review of the state of the art in various aspects of the Information systems discipline including Energy Informatics, Health IT, IT Business Alignment, IT Governance, Business Software trends, IT function in the global world and, finally, Big Data.

Given this multidisciplinary line, teaching approach must be aligned with that and lectures combine traditional classes, cases, mini-cases with invited talks with experts in the field covering very specific aspects and technologies. These experts are appointed to teach a class (around one hour) via Skype on a specific subject of their competence. On the other hand, the course responsible introduces the topic and wraps it up once the expert ends his or her lesson. Experts are coming from academia and industry. Calendar is given in advance to students and experts. In the academic year 2014-2015 four experts from three different countries were appointed for the course.

Assessment is manly based on two kinds of projects: individual and group. The first consists of a state of the art research project or a software implementation and customization of a system in one of the topics of the course. The weight of this individual project is 35% of the final grade. The second kind of project is developed in groups of 3-4 students and consists of a systematic literature review or a systematic mapping in one of the topics of the course or an advanced software development, customization or deployment covering one of the aspects of the course, typically Big Data aspects or advanced commercial information systems releases.

Big Data contents are specifically devoted to present main commercial technologies dealing with the capture, storage, analysis and visualization of big amounts of data. The approach is mainly based on the exploitation of these technologies in the energy sector, one of the cornerstones of the Østfold region and one of the main research areas at the Faculty of Computer Sciences at Østfold University College.

However, this is not the only course dealing with Big Data topics at master level. The "Machine Learning" course, deals with the topic from an analytics and purely computer science standpoint. After reviewing some of the most important methods on Machine Learning (Decision Trees, Neural Networks, Evolutionary Computation, etc.), the students gain practical knowledge to deal with challenges of data mining techniques (curse of dimensionality, missing values, overfitting, etc.) applied to both small and large datasets. A generalization of these big datasets can be considered Big Data given their volume, variety, velocity, value, variability, viscosity and virality. As stated before, the course is aimed to complement the aforementioned "Advanced Topics in Information Systems" course in order to provide students with the skills to implement specific algorithms in Big Data commercial solutions presented.

### 3.1.3 Continuous education

HiØ VIDERE is Østfold University College's center of competence that offers continuing education (with and without credits). HIØ VIDERE is a fixed administrative unit that offers over 50 different part-time studies and courses. The part-time studies and courses are offered online, in the evening, during the weekends and as seminars, and can be combined with work.

One of these part-time studies is "Big Data Analytics". The course is scheduled as a 15 ECTS credits track aimed to start in autumn 2015. The teaching approach includes lectures, tutorials, project work and individual study work. The course lasts one semester with four 2-day study collections in the daytime, a total of intensive 8 days scheduled to fit in lifelong learning and continuous education programmes in both public institutions and private organizations.

The course is a combination of hands-on mini projects and case studies with traditional lectures and demonstrations. The course aims to expose participants to some of the most recent ideas and techniques in Big Data Analytics. Starting with general concepts on Big Data´s, the course introduces the Datification concept, meaning capturing all quantifiable information. Data acquisition and collection processes are explained with practical examples, including collecting sensor data in smart grid, monitoring Twitter data in real-time, or capturing user behaviors on e-commerce websites. Relevant data mining and machine learning approaches are then provided and illustrated to show how data could be transformed into knowledge. The course focuses on scalable and popular models, which are being used in various well-known big data applications. Participants will learn how to do time-series analysis, with examples taken from consumption prediction in smart grid; or how to do sentiment analysis on real-time data collected from Twitter. The course will cover locality-sensitive hashing, which allows finding similar items in a large set of items. Many other large-scale algorithms are introduced as well, such as Stream Analysis, Nearest Neighbors, or Recommender Systems. Some popular machine learning models are also revisited, including Support Vector Machine, Random Forest, and clustering methods.

Privacy concerns are discussed in the final part of the course, which helps learners know their responsibilities for keeping personal information secure, including keeping data encrypted, and ensuring it is not kept for longer than necessary. Security solutions for these concerns are also introduced, which explains why building secure Big Data systems is so hard and surveys recent techniques that help; including learning direct processing on encrypted data, information flow control, and auditing.

By surveying state-of-the-art topics in Big Data, including data collection, storage, processing, and analytics; this course is providing learners with relevant tools to keep their companies stay competitive in new era or potentially define new business models in their industries. They will grasp emerging technologies and know how those technologies can be used to unlock value of data and effectively solve their business problems.

## 4. CONCLUSIONS AND FUTURE WORKS

Beyond the wars between causality and simple correlation of data, the Big Data phenomenon is here to stay and to grow. Apart from the obvious commercial lobbies from software vendors and integrators aimed to sell new solutions and services, it is a fact that datification will nurture as internet penetration, the internet of things and interactions grow. Taking out the byzantine discussions on the differences between traditional data-approaches and new big-data solutions, it is a fact that the overall panorama of data-science has evolved in a dramatic way in the last ten years. In this scenario the need of professionals with relevant skills to handle this complexity is not completely covered in the market. Higher education institutions must face this problem and react to market needs in a rapid and effective way. This efficiency leads to solutions in the short, mid and long term. Short-term solutions can be mapped to continuous education courses targeting professionals already working in the industry aimed to update their competences with specific courses on the topic. Mid-term solutions can be rooted on the design of specific master courses of undergraduates either fully devoted to Big Data aspects or containing courses on the topic. Finally, long-term solutions should be devoted to transform current courses refocussing them towards Big Data or, ideally and given the importance of the topic and the need for professionals, designing specific undergraduate programs on the topic or, finally, adopting the approach of specializations in bachelor programs (major and minor).

This paper presents the initiatives that are taking place at the Faculty of Computer Sciences, Østfold University College, Norway to integrate Big Data as a topic in its Bachelor, Master and continuous education courses. New master courses were implemented in the 2014-2015 academic year, while bachelor and continuous education courses will be deployed the year after. Initial results in terms of course assessments and students´ acceptation unveil promising perspectives for the initiative.

Future work will be twofold. Firstly, it is intended to assess the overall initiative in several aspects including student acceptation and industry acceptation, impact in research outputs and funded projects. Secondly it is intended to expand current initiatives in Big Data scope to more courses in the bachelor and master level and, finally, include Big Data topics as one of the cornerstones for research in the department including aspects as its management, capture, storage, analysis, visualization and governance.

## 5. ACKNOWLEDGMENTS

This research was supported by Østfold University College under project "Teknologi, samfunn og energi".